\documentclass[preprint]{revtex4-1}

\usepackage{amsmath, amssymb}
\usepackage{siunitx}
\usepackage{graphicx}   

\DeclareMathOperator\arctanh{arctanh}

\begin{document}

\title{Topological structures are consistently overestimated in functional complex networks}

\author{Massimiliano Zanin}
\affiliation{Centro de Tecnolog\'ia Biom\'edica, Universidad Polit\'ectica de Madrid, Madrid, Spain}
\affiliation{Universidade Nova de Lisboa, Lisboa, Portugal}

\author{Seddik Belkoura}
\affiliation{Innaxis Foundation \& Research Institute, Madrid, Spain}

\author{Javier Gomez} 
\author{Cesar Alfaro}
\author{Javier Cano}
\email{To whom correspondence should be addressed. E-mail: \href{mailto:javier.cano@urjc.es}{javier.cano@urjc.es}}
\affiliation{Universidad Rey Juan Carlos, Madrid, Spain}

\begin{abstract}
Functional complex networks have meant a pivotal change in the way we understand complex systems, being the most outstanding one the human brain. These networks have classically been reconstructed using a frequentist approach that, while simple, completely disregards the uncertainty that derives from data finiteness. We here provide an alternative solution based on Bayesian inference, with link weights treated as random variables described by probability distributions, from which ensembles of networks are sampled. By using both statistical and topological considerations, we prove that the role played by links' uncertainty is equivalent to the introduction of a random rewiring, whose omission leads to a consistent overestimation of topological structures. We further show that this bias is enhanced in short time series, suggesting the existence of a theoretical time resolution limit for obtaining reliable structures. We also propose a simple sampling process for correcting topological values obtained in frequentist networks. We finally validate these concepts through synthetic and real network examples, the latter representing the brain electrical activity of a group of people during a cognitive task. 
\end{abstract}


\date{This manuscript was compiled on \today}

\maketitle

Functional complex networks have brought an important advancement in the way complex systems are analysed. By shifting the focus from the underlying physical structures to the flow of information developing on top of them, functional networks yield a more detailed understanding of how, for instance, the human brain works \cite{bullmore2009complex, park2013structural}. 

The standard way of reconstructing such representations starts with the recording of a set of time series describing the dynamics of the nodes composing the system. In neuroscience, these typically reflect the evolution of physiological observables like electric (EEG) or magnetic (MEG) fields, or the consumption of oxygen by neurons (fMRI). Afterwards, the synchronous dynamics of pairs of nodes is assessed, using various metrics spanning from linear correlations to causalities \cite{bullmore2009complex, rubinov2010complex}. This approach is inherently {\it frequentist}, as a single value ({\it e.g.} the correlation coefficient) is extracted from each pair of nodes, and encoded as the weight of the corresponding link. Nevertheless, frequentist (or `classic') inference is not the only alternative, as proved by the long-standing controversy with Bayesian statisticians. For decades, researchers from both fields have fiercely defended the advantages of their respective approaches, with theoretical and practical evidence supporting the superiority of the Bayesian approach and of its axiomatic and decision theoretic foundations. 

The main conceptual difference between both approaches is that Bayesian inference considers data to be fixed, and the model parameters to be random, as opposed to what frequentist inference does. Furthermore, Bayesian inference---unlike frequentist---estimates a full probability model, including hypothesis testing. This entails several practical advantages, as: ({\it a}) incorporating prior knowledge about model parameters in a natural way; ({\it b}) accommodating any sample size, no matter how small; or ({\it c}) allowing more complex models, for which MCMC algorithms are guaranteed to converge, see \cite{robert2007tbc} for a more detailed discussion.

While Bayesian inference has previously been considered in neuroscience \cite{colombo2012bayes, hinne2012bayesian, janssen2014quantifying}, no attention has hitherto been devoted to the specific topic of functional network reconstruction.
Nevertheless, in the light of the different way data and parameters are treated within both frameworks, one question arises: do observed topological metrics vary, depending on which statistical approach (frequentist {\it vs.} Bayesian) is applied? 
We demonstrate here that this is actually the case by using both statistical and topological considerations. We further show how this bias implies that topological structures are consistently overestimated in the frequentist case, since the inherent uncertainty in the observed connectivity between nodes acts as a random rewiring process. We also prove that this bias is responsible for the existence of a minimum time resolution limit, below which no network structure can reliably be estimated; and provide an efficient algorithm to reduce it.

\section*{Frequentist vs.~Bayesian reconstruction of functional networks}\label{sec:reconstruction}

\begin{figure}[!tb]
\begin{center}
\includegraphics[width=0.60\textwidth]{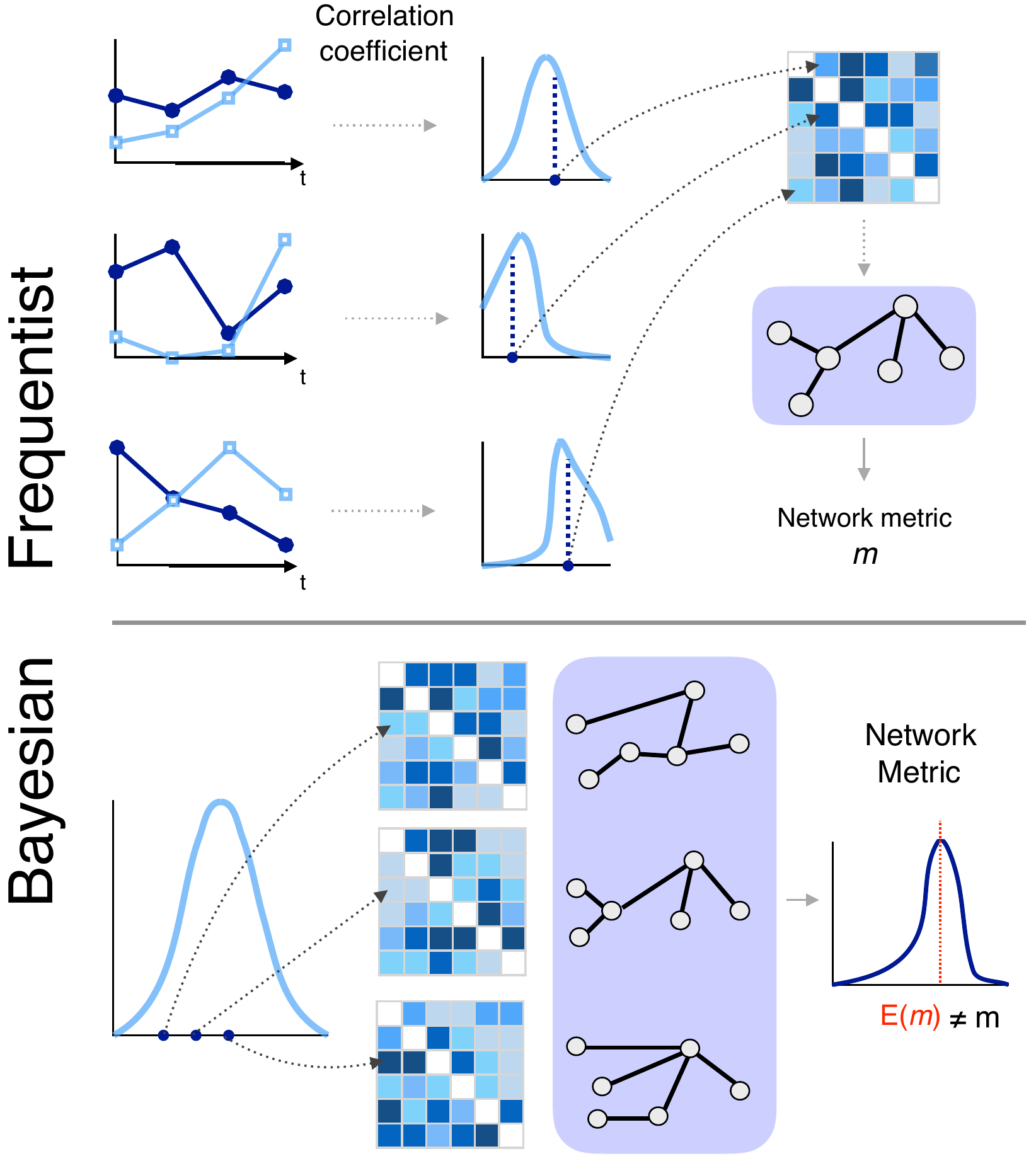}
\caption{Schematic representation of the functional network reconstruction process. The top part depicts the frequentist approach, in which the classical point correlation estimate is used for each pair of time series. The bottom part represents the Bayesian counterpart, in which several weight matrices are sampled from the correlation probability distributions.}\label{fig:Method}
\end{center}
\end{figure}

The standard procedure for network reconstruction is depicted in the upper part of Fig.~\ref{fig:Method}. 
The starting point of the process is a set of time series, describing the dynamics of the elements composing the system under study. Denoting by $X$ and $Y$ any two such series---assumed, without loss of generality, to come from a bivariate normal distribution---the frequentist approach assesses their connectivity through the Pearson's product-moment \emph{sample} correlation coefficient $r(X, Y)$ --- or its absolute value $|r(X, Y)|$. Note, however, that fixing the connectivity metric does not restrict the validity of results, see Discussion.
The correlation coefficients are afterwards mapped into an {\it adjacency matrix} $\mathcal{A}$ of size $N \times N$ ($N$ being the number of time series, hence of nodes). Such matrix is then usually pruned, in order to delete links of low statistical significance or weight, by applying a fixed threshold or by retaining a fixed fraction of the strongest links. Finally, a set of topological metrics is extracted from the resulting object. It is important to note that this approach implies that a point estimate $r$ is used to summarise the linear dependence between $X$ and $Y$. This hidden hypothesis is consistent with a frequentist approach, since it regards the correlation coefficient as a constant, estimating it through $r(X, Y)$.

On the other hand, the Bayesian approach puts a probability distribution over the \emph{population} correlation coefficient, $\rho$, and makes inference for it based on collected data. In doing so, we have assumed noninformative priors for all the involved parameters, see \cite{lee2012bsc} for details. Disregarding such probability distribution, which is a measure of the uncertainty in the connectivity, is readily expected to introduce biases in the obtained results. In general terms, the extraction of a metric $m$ can be seen as the application of a highly complex and non-linear function of the adjacency matrix, $m = f(\mathcal{A})$. When it comes to compute the expected value of the function of a given random variable $X$, it is well-known that, in general, $E[f(X)]  \neq f(E[X])$, being the latter the wrong way to do it. Specifically, if one has a sample $x = (x_1, x_2, \ldots, x_n)$, the right method to compute $E[f(x)]$ implies evaluating $f$ for all the elements in the sample, {\it i.e.} $(f(x_1), f(x_2), \ldots, f(x_n))$, and finally averaging such values. The incorrect way would calculate the sample mean $\bar{x} = E[x]$ first, and evaluate $f(\bar{x})$ afterwards. 
As depicted in the bottom part of Fig.~\ref{fig:Method}, the correct procedure for reconstructing functional networks thus entails: (1) sampling different values of the link weights, according to the posterior distribution of $\rho|data$; (2) creating multiple networks, one for each sampled weight set; (3) computing the corresponding target metric $m$ for each network; (4) obtaining the empirical probability distribution of the target metric $m$; and (5) calculating its expected value  $E[m]$.

\begin{figure}[!tb]
\begin{center}
\includegraphics[width=0.60\textwidth]{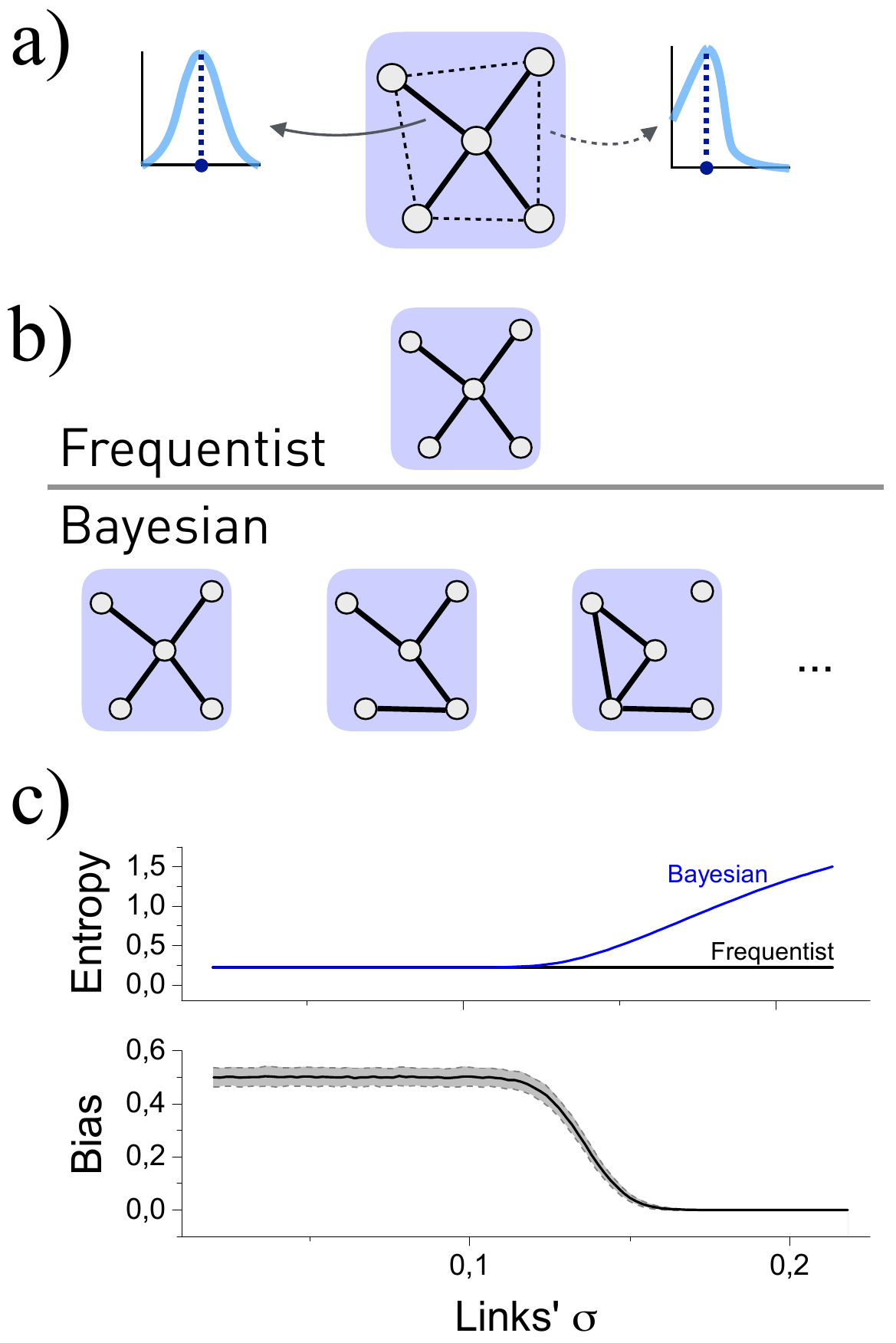}
\caption{Example of the analysis of a star-like structure. {\it a}) Initial structure, with four nodes strongly connected with a central one, and loosely connected between them. {\it b}) Frequentist and Bayesian results. In the former case the output is a constant network, as uncertainty is disregarded; in the latter, and due to the inherent uncertainty, different structures are generated, some of them different from the original one. {\it c}) Evolution of the entropy and of the bias (see Appendix \ref{app:bias} for definitions) as a function of the links' uncertainty $\sigma$.}\label{fig:Star}
\end{center}
\end{figure}

Beyond this statistical consideration, the error made by the frequentist approach can also be understood from a topological point of view. Let us suppose one is analysing a star-like network, as depicted in Fig.~\ref{fig:Star} {\it a}). As the strongest links are those connected with the central hub, and since the frequentist approach disregards their associated uncertainty, the result of the reconstruction process would always be constant: a well-defined star-like structure ---see Fig.~\ref{fig:Star} {\it b}) Top. On the other hand, the Bayesian approach recognises that silent links actually have non-null weights---as described by their posterior probability distribution $\rho|data$---albeit with lower expected values than active ones. When link weights are eventually sampled from the corresponding distribution, links between peripheral nodes may actually have greater weights than the central ones.
The result is a set of networks in which the star-like structure is sometimes lost---Fig.~\ref{fig:Star} {\it b}) Bottom. If one then analyses the resulting structure---through {\it e.g.}~the entropy of the degree distribution---two completely different results are found: a low constant entropy in the frequentist case, and a variable and higher value in the Bayesian case, see Fig.~\ref{fig:Star} {\it c}), and Appendix \ref{app:metrics} for metric definitions.

Both Figs.~\ref{fig:Method} and \ref{fig:Star} suggest two important conclusions. First, that disregarding the inherent uncertainty associated with functional links introduces a bias in the obtained topological metrics. Secondly, that the link uncertainty acts like a random {\it rewiring process}, such that dismissing it overestimates the regularities observed in the network.

\section*{Application to brain physiological data}
\label{sec:application_brain_data}

In order to illustrate how the previously defined bias may affect the analysis of real-world networks, we consider here a large set of functional networks representing brain dynamics in control and alcoholic subjects---see Appendix \ref{app:EEG} for details on the data set.

\begin{figure*}[!tb]
\begin{center}
\includegraphics[width=0.99\textwidth]{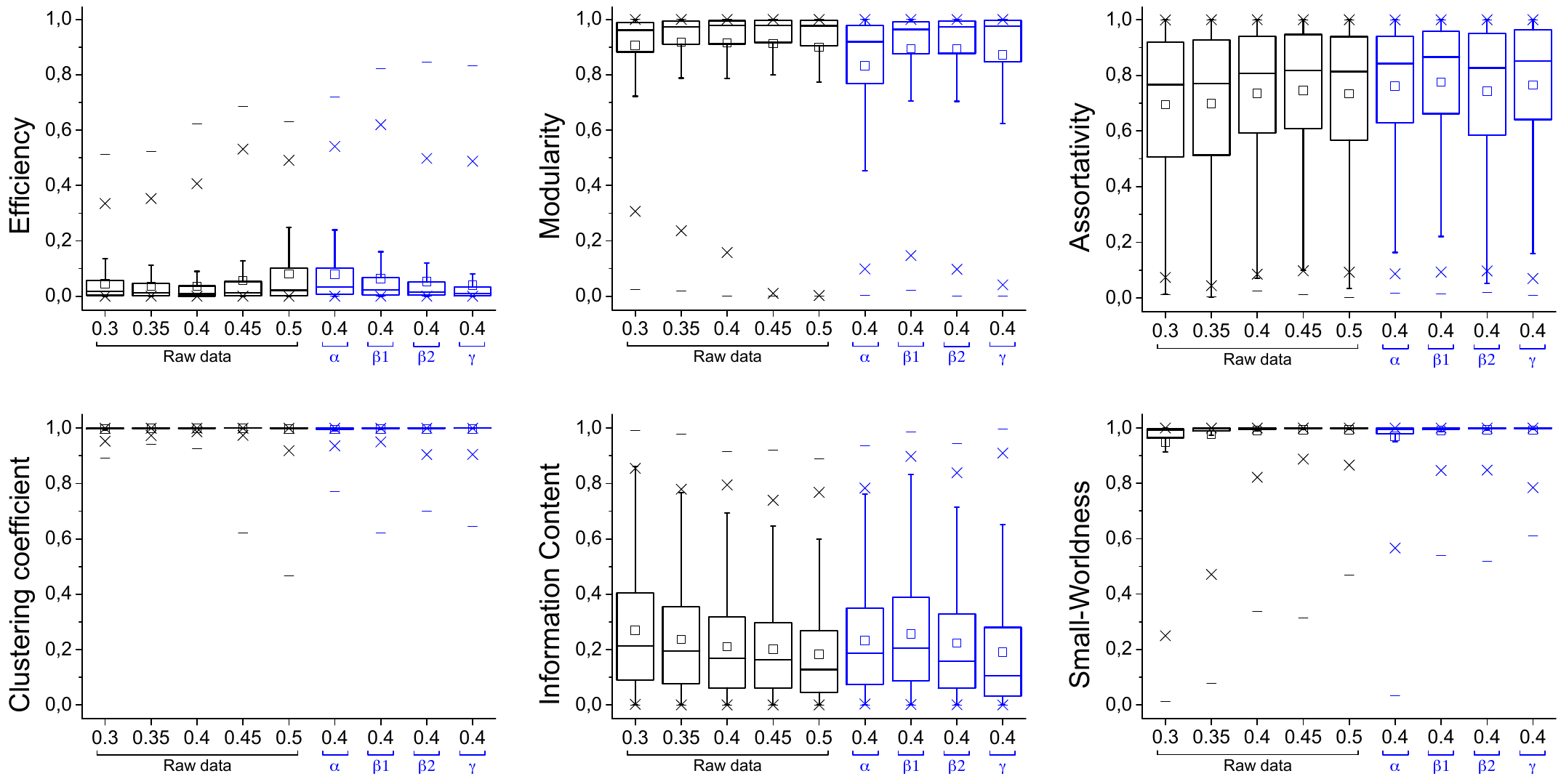}
\caption{Distribution of the bias between frequentist and Bayesian values in real EEG networks, for the six considered topological metrics---see Appendices \ref{app:metrics}-\ref{app:EEG} for definitions and experimental data description. Each box plot corresponds to networks obtained with a fixed link density (from $0.3$ to $0.5$) and filtered by four frequency bands. Central horizontal bars and squares represent the median and the mean of the distribution, respectively; boxes and crosses the 25th-75th and 1th-99th percentiles; and the external horizontal lines the minimum and maximum.}\label{fig:MainRes}
\end{center}
\end{figure*}

Fig.~\ref{fig:MainRes} depicts the distribution of the bias observed in six metrics commonly used in complex network, see Appendix \ref{app:metrics}. Networks have been reconstructed using different criteria, including five different link densities in the binarisation process, and five band filtering ({\it i.e.} the raw time series, and bands $\alpha$, $\beta 1$, $\beta 2$, and $\gamma$). It can be appreciated that the frequentist value overestimate the metrics in all cases, except for the Efficiency \cite{latora2001efficient} and the Information Content \cite{zanin2014information}. However, this is consistent with the insights of Fig.~\ref{fig:Star}, which indicates that the rewiring introduced by the Bayesian method implies that all topological metrics are overestimated by the frequentist approach. The two exceptions---Efficiency and Information Content---are explained by the fact that such metrics are actually maximal for random networks.

Of special relevance is the analysis of the behaviour of the Small-Worldness, a metric assessing the coexistence of a high number of triangles with short geodesic distances \cite{humphries2008network}. In spite of some critical voices \cite{muller2014brain, papo2016beware}, Small-Worldness has been considered as one of the landmarks of brain functional networks, describing their capacity for the simultaneous local integration and long-range transmission of information \cite{bassett2006small,bassett2016small}. Beyond physiological \cite{papo2016beware} or methodological \cite{hlinka2017small} reasons that may bias the observed Small-Worldness, we have shown that this property is also affected by the use of a frequentist approach. Fig.~\ref{fig:MainRes} suggests that the small-world nature of the human brain should be taken with caution: even if the brain seems to possess such feature, the actual value may have been substantially overestimated.

\section*{Time series length}
\label{sec:TSL}

The selection of the optimal time series length for reconstructing functional networks is, in general, a non-trivial problem, particularly complex in neuroscience. If, on one hand, long time series may seem desirable for a better estimation of functional connectivity, this should be balanced, on the other hand, by the need of a stationary dynamics. In other words, if a given cognitive task is executed in around one second, this is the maximum length that can be considered without introducing spurious information.
This issue has recently been studied in, for instance, \cite{fraschini2016effect}, finding that the time series length has profound effects in the observed topological metrics.

On top of any physiological consideration, we here note that shorter time series imply higher uncertainty in the estimation of the connectivity metric---the correlation coefficient in our case.
This can be better understood by taking the star-like structure of Fig.~\ref{fig:Star} as an example.
Suppose that the real structure driving the system's dynamics has a star-like topology; and that, when calculated using the frequentist approach, the resulting functional network matches exactly the real one. By acknowledging the inherent randomness of the links' weights, the Bayesian approach would suggest that the star-like structure is not the only possible one, but (eventually, given small enough uncertainties) just the most probable one.
In other words, if the length of the used time series is not enough to ensure a small uncertainty in the links' $\sigma$, results yielded by the frequentist approach cannot be trusted, even if actually correct.

This issue is studied in Fig.~\ref{fig:ResSM}, which reports the results obtained with a synthetic model---see Appendix \ref{app:synth} for a description. The two panels depict the evolution of the fraction of functional links wrongly observed in the reconstructed networks, as compared to the true connectivity, as a function of the number of data points and the coupling strength. The top blue region of both panels indicates that high couplings lead to an over-synchronisation of the system, and thus to the observation of a spurious all-to-all connectivity.
A more interesting behaviour emerges for intermediate couplings ($\gamma \approx 0.5$): while both methods converge towards the correct topology, the Bayesian  approach requires substantially longer time series to reach the same precision.
This result tells us that it is possible for the frequentist approach to detect the real functional structure, provided the information encoded in the data is explicit enough---as it has been shown in this tailored example. On the other hand, the Bayesian approach requires longer time series to resolve the topology, {\it i.e.} to reduce the weight uncertainty enough to reach a stable structure.

\begin{figure}[!tb]
\begin{center}
\includegraphics[width=0.65\textwidth]{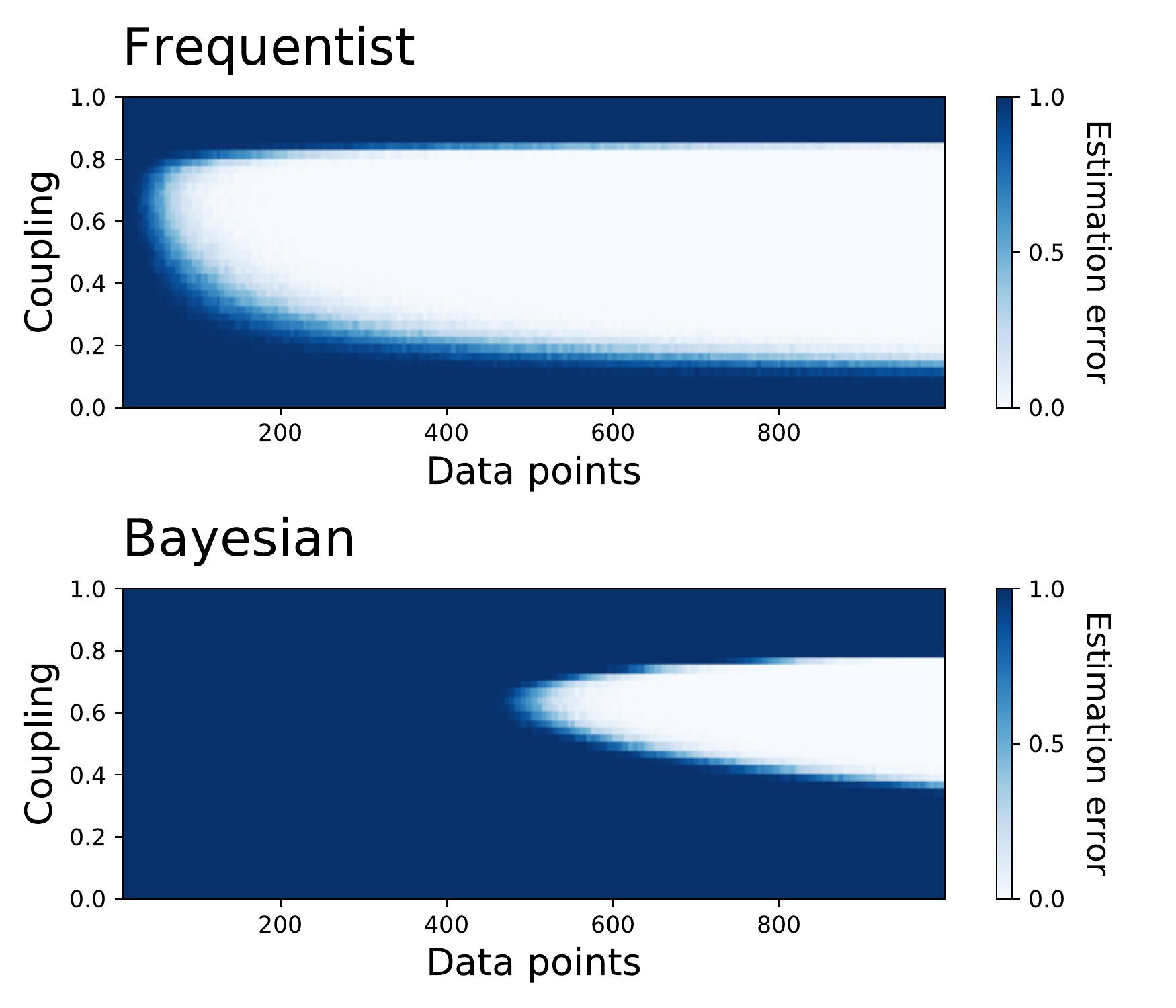}
\caption{Impact of the time series length in the topological uncertainty. Evolution of the structural estimation error as a function of the coupling constant $\gamma$ and the time series length, for the frequentist (Top) and Bayesian (Bottom) approaches, using a synthetic functional model---see Appendix \ref{app:synth} for definitions.}\label{fig:ResSM}
\end{center}
\end{figure}

Summarising, the Bayesian approach always reminds that different (real) alternative connectivity patterns could yield time series that are compatible with the observed frequentist functional connectivity. In order to resolve this multiplicity in solutions, longer time series---implying a reduction in the link weight uncertainty---should be considered. Frequentist results should therefore not be taken at face value, especially in real-world analyses, as there are uncountable, {\it a priori} unknown, situations where they may just be stemming from biases.

\section*{Correcting frequentist networks through rewiring}
\label{sec:Correcting}

The Bayesian approach yields a more complete and theoretically correct view of the system under study; this, however, comes at some costs. First, a Bayesian version may not be available for many connectivity metrics. But even if so, the associated computational burden may be prohibitive in large-scale studies. We provide here an affordable alternative, based on the creation of a set of rewired networks simulating the Bayesian output.

Let us start with the case when a complex network has been obtained using the frequentist methodology, {\it i.e.}~when an adjacency matrix as the one in Fig.~\ref{fig:Method} Top has been calculated. Instead of using the fully Bayesian approach to obtain the probability distribution associated with each link, we propose here an alternative low-cost and efficient procedure, based on the Fisher's  transformation of the correlation coefficient, $Z(\rho) = \arctanh (\rho)$, see \cite{fisher1915frequency,fisher1921probable}. It is well-known that $Z(\rho)$ follows approximately a normal distribution with standard deviation $\sigma_Z =
1 / \sqrt{df - 3}$, where $df$ is the effective number of degrees of freedom, which coincides with the series length $n$ if data are independent. In case of autocorrelated time series, as the ones considered here, an effective number of degrees of freedom has to be defined as:
\begin{equation*}
\frac{1}{df} \approx \frac{1}{n} + \frac{2}{n}\sum_{\tau}\varrho_{ii}(\tau)\varrho_{jj}(\tau),
\end{equation*}
where $\varrho_{xx}(\tau)$ is the autocorrelation of signal $x$ at lag $\tau$, see \cite{valencia2009complex} for details. We can then invert the transformation, and assume that $\rho$ can be reasonably described through  a normal distribution $\mathcal{N}[r, \tanh( \sigma_Z ) ]$. 
Afterwards, an ensemble of synthetic networks is created, sampling the weight of each link from the corresponding distribution, and applying a threshold to recover a network with the same link density as the original frequentist one. Provided this approximation is good enough, calculating topological metrics on this ensemble is equivalent to computing them on an set of networks created using the Bayesian approach. Additionally, this approach can be applied with any connectivity metric, provided it can be described by a known probability function.

\begin{figure*}[!tb]
\begin{center}
\includegraphics[width=0.99\textwidth]{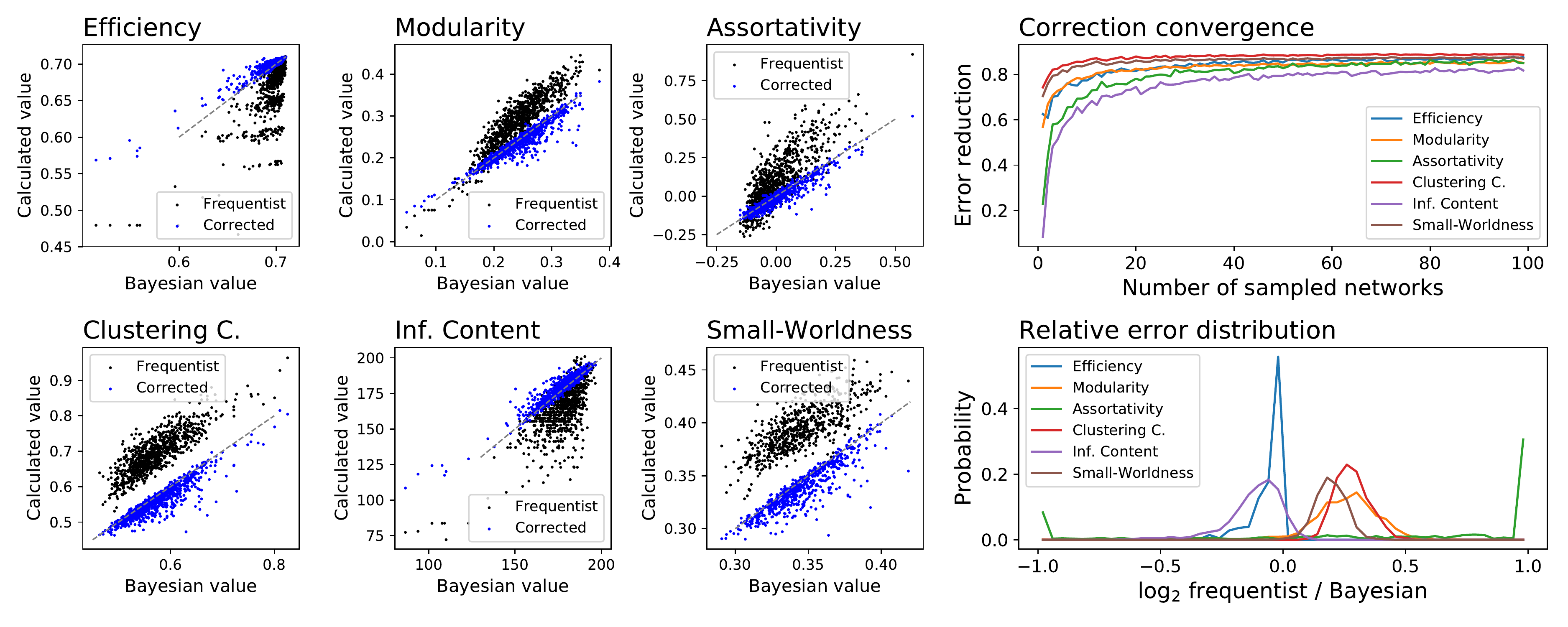}
\caption{(Left panels) Comparison of Bayesian, frequentist and corrected frequentist topological values (see main text for a definition of the latter), for the EEG data set described in Appendix \ref{app:EEG}. (Right top panel) Evolution of the reduction in the error, defined as the fraction of the error incurred by the frequentist approach disappearing after the synthetic correction, as a function of the number of sampled synthetic networks. (Right bottom panel) Probability distribution of the relative error associated to the frequentist approach, defined as the $\log_2$ of the ratio between the values of the frequentist and Bayesian topological metrics. }\label{fig:ResCorr}
\end{center}
\end{figure*}

To demonstrate the effectiveness of this correction method, Fig.~\ref{fig:ResCorr} reports six scatter plots, one for each considered topological metric, comparing the Bayesian, frequentist and corrected frequentist values for the same networks analysed in Fig.~\ref{fig:MainRes}. It can be appreciated that the latter is an excellent approximation of the Bayesian process, while still saving orders of magnitude of computational cost. Additionally, Fig.~\ref{fig:ResCorr} Top Right shows the evolution of the fraction of the recovered error as a function of the number of drawn synthetic networks. As we can observe, the frequentist bias can be reduced by around $80\%$ with as low as $100$ realisations.

\section*{Discussion}
\label{sec:discussion}

The fact that the existence and strength of links in functional networks cannot assuredly be defined has profound implications in the topological analysis. As opposed to the classical frequentist point of view, we  presented here a Bayesian approach, and demonstrated its theoretical advantages and capacity to account for the weights' inherent uncertainty.
We have shown that, in general, reconstructing functional networks using the frequentist methodology overestimates the presence of regularities and non-trivial ({\it i.e.} non-random) structures. As shown in Fig.~\ref{fig:ResCorr} Bottom Right, the Clustering Coefficient and the Modularity were overestimated, on average, by $19\%$, and the Small-Worldness by  $13\%$. We further proved that such drawback is aggravated by the use of short time series, although it is possible to (partially) correct it by sampling synthetic networks.

Such topological bias is a general phenomenon, independent on the actual synchronisation or connectivity metric used, and on the data set considered. Data cannot represent the whole universe, and when coming from real observations they are usually polluted by noise. Therefore, any metric based on them is inherently uncertain and fuzzy, and topological biases, as the one we have shown here, will always appear to a greater or lesser degree. This effect is to be expected in the analysis of any real-world system in which functional representations are relevant, as {\it e.g.} financial markets \cite{bonanno2003topology}, medicine \cite{barabasi2011network}, or companies' \cite{johannissson1998personal} and social networks \cite{cross2004hidden}. 

If results presented in all research works analysing functional networks are potentially biased, this does not mean their conclusions are {\it de facto} wrong. For instance, while the Small-Worldness of brain functional networks may have been overestimated, to such a point that their actual value cannot be trusted, the existence of a positive value still suggests that the brain has a small-world structure---even if less marked. Additionally, functional networks corresponding, for instance, to different diseases, can still be compared, provided the respective uncertainties (and hence, the time series length) are similar.

\appendix

\section{Topological metrics}
\label{app:metrics}

For the sake of completeness, we briefly describe here the topological metrics considered in this work. For more thorough definitions, the reader can consult the corresponding references, or the many reviews available in the literature \cite{boccaletti2006complex, costa2007characterization}.

\begin{enumerate}

\item {\it Efficiency}. Measure of how efficiently information can be transmitted in a network, and defined as the inverse of the harmonic mean of the geodesic distance between nodes \cite{latora2001efficient}:
\begin{equation}
E = \frac{1}{n(n-1)}\sum_{i, j\neq i}\frac{1}{d_{i,j}}.
\end{equation}

\item {\it Modularity}. Presence of communities, {\it i.e.} groups of nodes more connected between them than with the remainder of the network \cite{fortunato2010community}.

\item {\it Assortativity}. Conditional probability $P(k'|k)$ that a link from a node of degree $k$ points to a node of degree $k'$. It is calculated as the Pearson correlation coefficient of the degrees at either ends of a link.

\item {\it Clustering coefficient}. Measure of the presence of triangles in the network, calculated as the relationship between the number of triangles in the network and the number of connected triples. 

\item {\it Information Content}. Measure assessing the presence of meso-scale structures in complex networks, based on the identification of regular patterns in the adjacency matrix of the network, and on the calculation of the quantity of information lost when pairs of nodes are iteratively merged \cite{zanin2014information}.

\item {\it Small-Worldness}. Metric capturing the degree of {\it Small-Worldness} of a network, defined as the coexistence of a high Clustering Coefficient and a low mean geodesic distance \cite{humphries2008network}.

\item {\it Entropy of the degree distribution}. Metric measuring the heterogeneity, in terms of Shannon's entropy, of the distribution created by nodes' degree \cite{wang2006entropy}:
\begin{equation}
H = - \sum _k p(k) \log_2 p(k).
\end{equation}
The minimum $H = 0$ indicates a constant degree across all nodes, while higher values a more uniform distribution of degrees.

\end{enumerate}

\section{EEG brain recordings}
\label{app:EEG}

As an example of application of the proposed methodology, we considered a data set of EEG recordings from a group of alcoholic subjects and matched controls \cite{zhang1995event, cao2014disturbed}, freely available at \url{https://archive.ics.uci.edu/ml/datasets/EEG+Database}. Each trial corresponds to an object recognition task, as described in \cite{snodgrass1980standardized}; and its corresponding EEG activity has been recorded during one second, with a $256$Hz ($3.9$-ms/ epoch) sampling rate from $64$ electrodes located at standard scalp sites. A total of $900$ trials were analysed, half of them from control subjects, the remainder from alcoholic.
Along with the raw time series, different filtering were considered, corresponding to the bands $\alpha$ ($8.0$ - $13.0$ Hz), $\beta 1$ ($13.0$ - $20.0$ Hz), $\beta 2$  ($20.0$ - $30.0$ Hz) and $\gamma$  ($30.0$ - $50.0$ Hz).

\section{Bias calculation}
\label{app:bias}

For each set of time series, the frequentist method entails calculating one single functional network, and, from it, a (single valued) topological metric. On the other hand, the Bayesian approach entails sampling a large number of networks, each one with a potentially different topology, from which we extract the probability distribution associated with the topological metric. Specifically, we have used $10^5$ Bayesian networks in our experiments, thus yielding $10^5$ metric values.

The bias introduced by the frequentist approach is then evaluated by calculating the proportion of sampled (Bayesian) values that are smaller than the frequentist one. Note that a bias of $0.5$ indicates that the frequentist value coincide with the median of the Bayesian distribution, and thus that both approaches are equivalent; while values close to $0.0$ or $1.0$ indicate a frequentist under- and overestimation of the metric, respectively.

\section{Synthetic functional network model}
\label{app:synth}

We simulate a system composed of $n = 10$ elements, connected according to a star-like structure, whose adjacency matrix is:

\begin{equation}
\mathcal{A} = 
\begin{bmatrix}
   0 & 1 & \ldots & 1 \\
   1 & \ddots & & \\
   \vdots & & \text{\Huge0} & \\
   1 & & & \ddots 
   \end{bmatrix}
\end{equation}

The output of each element is initially a random number sampled from a normal distribution $\mathcal{N}(0, 1)$. Afterwards, such numbers are coupled according to:

\begin{equation}
x_i(t+1) = (1 - \gamma) x_i(t) + \frac{\gamma}{\sum _{j \neq i} a_{j, i}} \sum _{j \neq i} a_{j, i} x_j(t),
\end{equation}

where $0 \leq \gamma \leq 1$ is the coupling constant.
The resulting time series are then used to reconstruct the observed functional network $\mathcal{F}$; and the relative error between $\mathcal{F}$ and the original $\mathcal{A}$ is defined as:

\begin{equation}
e = \frac{1}{n(n-1)} \sum _{i, j} | a_{i,j} - f_{i, j} | .
\end{equation}

%

\end{document}